\documentclass{optica-article}

\articletype{Research Article}
\makeatletter
\def\@journalname{Optica}
\makeatother

\usepackage{epstopdf}
\usepackage{lineno}
\usepackage{amsmath}
\usepackage{bm}
\usepackage{graphicx}

\begin{document}

\title{Full-Vector Diffractive Deep Neural Networks}

\author{Shengmei Zhao,\authormark{1,2,*} Gongyuan Wang,\authormark{1} Zhiwan Wang,\authormark{1} Jianan Shao,\authormark{1} and Haolin Liu\authormark{1}}

\address{\authormark{1}Institute of Signal and Information Processing and Transmission, Nanjing University of Posts and Telecommunications(NUPT), Nanjing 210003, China\\
\authormark{2}Key Lab of Broadband Wireless Communication and Sensor Network Technology, NUPT,Ministry of Education, Nanjing 210003, China}

\email{\authormark{*}zhaosm@njupt.edu.cn} 


\begin{abstract*}
Conventional diffractive deep neural networks (D$^2$NNs) treat polarization as independent channels and rely on scalar or semi-vectorial propagation models, which neglect cross-polarization coupling and vector diffraction. Here we propose a vector D$^2$NN (V-D$^2$NN) that embeds the full-vector angular spectrum method into the end-to-end training pipeline. This physically rigorous framework describes vectorial light–matter interactions across cascaded diffractive layers, enabling direct optimization of polarization conversion, spin–orbit coupling, and vectorial interference without external polarization optics. We demonstrate the V-D$^2$NN on polarization-multiplexed tasks—vector beam generation, polarization-dependent imaging and classification, and multi-channel optical encryption—where it consistently outperforms scalar and semi-vectorial counterparts. In longitudinal-field engineering, the V-D$^2$NN actively shapes a prescribed longitudinal field with a normalized correlation of 0.839, a capability inaccessible to scalar-propagation models. An open-source training framework is also provided to support further development of vectorial diffractive optics for computing, sensing, and communications.
\end{abstract*}

\section{Introduction}

Diffractive deep neural networks (D$^2$NNs) implement complex optical field transformations through cascaded diffractive layers, providing an all-optical computing platform that operates at the speed of light \cite{Lin2018}. These passive systems exploit the inherent parallelism and bandwidth of free-space propagation without active components, circumventing the latency and energy costs of electronic processing \cite{Veli2021, Li2024AllOptical}. D$^2$NNs have been applied to image classification, wavefront shaping, spatial-mode multiplexing, structured light generation, and other tasks \cite{Chen2021, Feng2021, Sun2021, Zhang2021, Hu2026, Xiao2026}.

The polarization degree of freedom provides a natural route to higher channel capacity and information density, driving recent interest in polarization-aware D$^2$NNs \cite{Rong2025, Wang2024}. Reported strategies include training polarization-insensitive neurons for orthogonal channels \cite{Luo2022Polarization}, inserting non-trainable polarizer arrays between layers \cite{Li2022}, employing anisotropic meta-atoms for polarized D$^2$NNs (P-D$^2$NNs) \cite{Zhang2022}, and integrating Jones-matrix metasurfaces (M-D$^2$NNs) \cite{Wang2024}. These approaches have produced functional devices, but they share a common approximation: polarization is treated as an independent channel attribute rather than an intrinsic property of a vectorial wave. Propagation is described by scalar or semi-vectorial models that neglect cross-polarization coupling and vector diffraction—effects that become significant in high-numerical-aperture systems or with complex meta-atoms \cite{Mueller2017}. Consequently, such networks cannot naturally capture polarization conversion, spin–orbit coupling, or vectorial interference, leaving much of the potential of vectorial optics unexploited.

In parallel, several refinements to the angular spectrum method have been proposed to improve its accuracy for vectorial diffraction problems. Mansuripur introduced a \(3\times 2\) transition matrix that maps the incident transverse polarization components to the full diffracted field, including the longitudinal component \cite{Mansuripur1989}. This approach, while useful for certain vector diffraction problems \cite{Yeh1998}, does not strictly satisfy the vector Helmholtz equation or the transversality condition. Subsequent work by Davis employed a vector potential formulation \cite{Davis1979}, and later, Hertz vector approximations were developed to more accurately describe Gaussian beam propagation \cite{Gillen2009}. More recently, Martínez-Herrero et al. proposed an integral angular spectrum method that decomposes the incident field into s- and p-polarized components \cite{MartinezHerrero2013}. These methods improve upon the standard angular spectrum formulation, yet they often encounter limitations in high-numerical-aperture cases, lack a closed-form fast-Fourier-transform implementation, or neglect the longitudinal polarization component and the effects of dielectric interfaces. Nevertheless, recent advances in scalable angular spectrum propagation \cite{Heintzmann2023} and angular spectrum rearrangement \cite{Hao2025} have extended the applicability of these methods, while the generic full-vector angular spectrum method \cite{Song2025} offers a more comprehensive framework for calculating diffraction of arbitrary electromagnetic fields. 

In the paper, We propose a fundamentally different approach, the full-Vector Diffractive Deep Neural Network (V-D$^2$NN), which treats the optical field as a genuine vectorial quantity from the outset. Unlike scalar or semi-vectorial models that approximate polarization as an independent channel, V-D$^2$NN explicitly accounts for the vector nature of light throughout propagation and modulation. The central idea is to embed the full-vector angular spectrum method directly into the differentiable training pipeline. This enables accurate modeling of cross-polarization coupling and vector diffraction during free-space propagation, while modulation at each diffractive layer is described as a vectorial transformation that naturally captures polarization conversion and anisotropic responses. The network can thus learn to exploit spin–orbit interactions and polarization-structured fields as intrinsic computational resources. By incorporating vectorially accurate propagation, V-D$^2$NN can address tasks that remain inaccessible to scalar or semi-vectorial networks. We demonstrate its performance on several polarization-multiplexed problems—vector beam generation, polarization-dependent imaging, and multi-channel optical encryption—where it consistently offers improved efficiency and design flexibility compared with scalar and semi-vectorial counterparts. The architecture supports full vectorial propagation and polarization manipulation within a single diffractive network, with a propagation model that explicitly accounts for cross-polarization effects in cascaded layers. An open-source training framework is also provided to facilitate further exploration in optical computing, sensing, and communications.

The remainder of the paper is organized as follows. Section~2 describes the theoretical basis of V-D$^2$NN. Section~3 presents simulation and experimental results. Section~4 discusses advantages and potential applications, and Section~5 concludes.

\section{Full-Vector Diffractive Deep Neural Networks}
\begin{figure}[!htpb]
    \centering
    \includegraphics[width=0.8\linewidth]{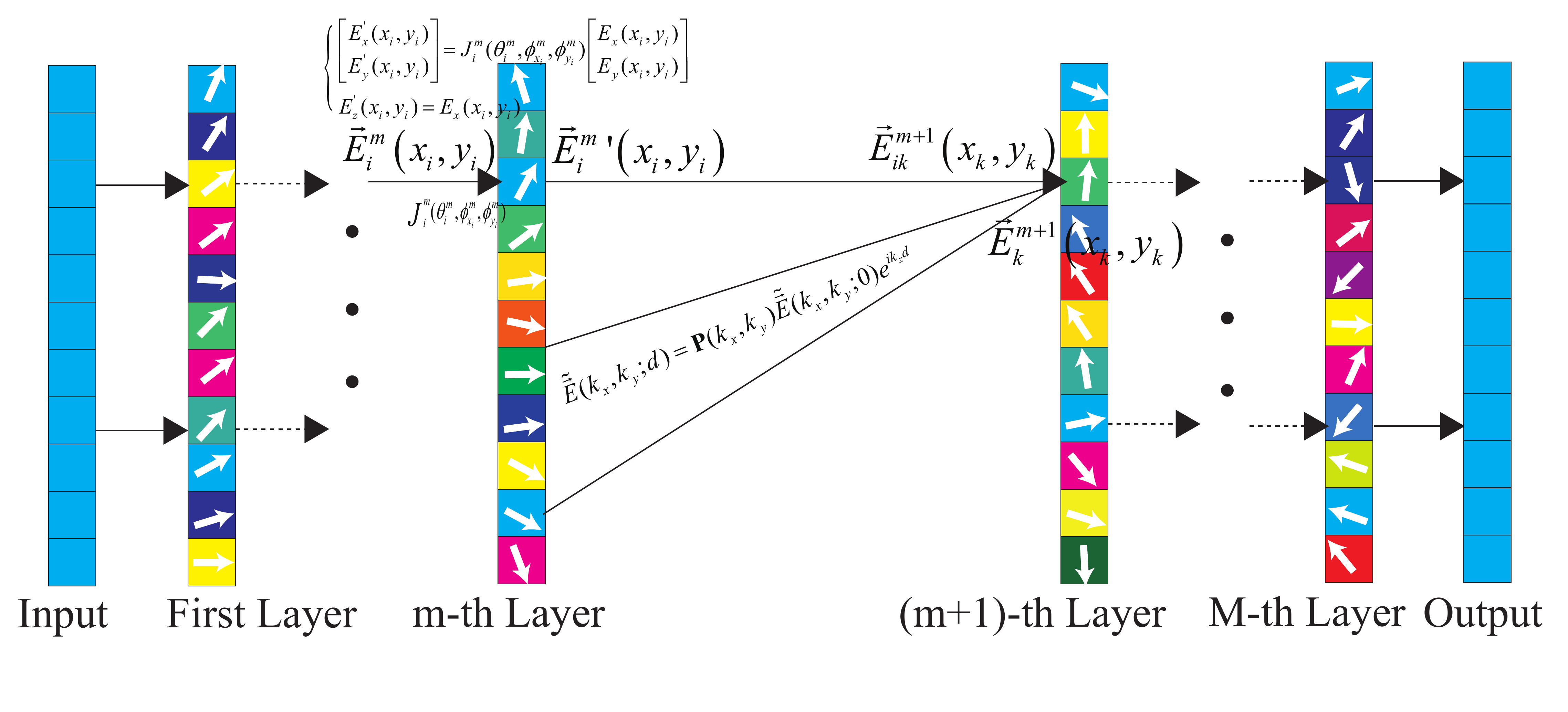}\vspace{-5mm}
    \caption{The schematic diagram of the proposed V-D$^2$NN.}

    \label{fig:Fig1}
\end{figure}

The V-D$^2$NN architecture is shown in Fig.~\ref{fig:Fig1}. It consists of multiple diffractive layers, transmissive or reflective, separated by fixed propagation distances. The first and last layers define the input and output planes, while the intermediate layers act as hidden layers. Each pixel on a layer functions as a vectorial transformation, and each neuron is fully connected to all neurons in the subsequent layer, following the vector diffraction theory. In forward propagation, the field emitted from one layer serves as the input to the next. 

For the \(m\)-th layer, the \(i\)-th neuron located at \((x_i,y_i)\) receives an incident vector field \(\vec{E}^m_{i}(x_i,y_i)\). The neuron modulates this field through a complex transmission or reflection coefficient, represented here by a Jones matrix \(\mathbf{J}^m_i(x_i,y_i)\) acting on the transverse polarization components. The modulated field then propagates to the next layer and contributes to the field incident on it.

A complete description of the field at the input plane requires a three-component complex electric field:
\begin{equation}
\vec{E}^m_{i}(x_i,y_i)=
\begin{bmatrix}
E_x(x_i,y_i)\\
E_y(x_i,y_i)\\
E_z(x_i,y_i)
\end{bmatrix}.
\end{equation}
Such a description is physically necessary, as it allows the field to satisfy both the vector Helmholtz equation and the transversality condition:
\begin{align}
\nabla\times\nabla\times \vec{E} - k^2\vec{E} &= \vec{0}, \\
\nabla \cdot \vec{E} &= 0.
\end{align}
In the frequency domain, the transversality condition translates to the familiar orthogonality constraint on each plane-wave component:
\begin{equation}
\vec{k} \cdot \tilde{\vec{E}}(\vec{k}) = 0, \qquad \vec{k} = (k_x, k_y, k_z)^T,
\end{equation}
which requires the electric-field spectrum to remain perpendicular to its propagation direction. As a result, even an initially transverse field acquires a longitudinal component during propagation. This effect is absent in the standard two-component Jones formalism, and therefore a three-component approach is essential for accurately describing polarization evolution and component coupling in full vectorial propagation.

In this architecture, polarization-specific tasks are encoded at the input layer, processed through the hidden layers, and the output layer is configured according to the target task. The field evolves from one layer to the next through two successive operations: \textbf{vector modulation} at the diffractive layer, followed by \textbf{full-vector propagation} in free space. Each neuron is fully connected to all neurons in the following layer, consistent with vector diffraction theory. Regardless of whether the input is specified as a scalar field, a Jones vector, or a three-component field, the network maintains a unified three-component vector representation throughout propagation and optimization, ensuring that the mapping from the input plane to the output plane is treated consistently as a vectorial transformation.

\subsection{Vector Modulation at Diffractive Layers}

 The modulation at each layer is described by a position-dependent \(2\times2\) Jones matrix \(\mathbf{J}^{m}_i(x_i,y_i)\) acting on the transverse field components. The longitudinal component is left unchanged: \(E'_z(x_i,y_i)=E_z(x_i,y_i)\). This matrix encodes a general anisotropic response, parameterized by three trainable parameters per pixel: the rotation angle \(\theta_i^{m}\) of the local principal axes, and two phase retardances \(\phi_{x_i}^{m}\), \(\phi_{y_i}^{m}\) along those axes. The transverse modulation is written as
\begin{eqnarray}
\begin{bmatrix}
E'_x(x_i,y_i)\\
E'_y(x_i,y_i)
\end{bmatrix}
&=&
\mathbf{J_i}^{m}\!\left(\theta_i^{m},\phi_{x_i}^{m},\phi_{y_i}^{m}\right)
\begin{bmatrix}
E_x(x_i,y_i)\\
E_y(x_i,y_i)
\end{bmatrix},
\\
E'_z(x_i,y_i)&=&E_z(x_i,y_i),
\end{eqnarray}
where \(\theta_i^{m}\) defines the orientation of the principal axes, and \(\phi_{x_i}^{m}\), \(\phi_{y_i}^{m}\) are the phase shifts along them. The Jones matrix adopts the standard rotation–phase–rotation form:
\begin{equation}
\mathbf{J_i}^{m}(x_i,y_i)
=
\mathbf{R}\!\left(\theta_i^{m}(x_i,y_i)\right)
\begin{bmatrix}
e^{i\phi_{x_i}^{m}(x_i,y_i)} & 0\\
0 & e^{i\phi_{y_i}^{m}(x_i,y_i)}
\end{bmatrix}
\mathbf{R}\!\left(-\theta_i^{m}(x_i,y_i)\right),
\end{equation}
with
\begin{equation}
\mathbf{R}(\theta)=
\begin{bmatrix}
\cos\theta & -\sin\theta\\
\sin\theta & \cos\theta
\end{bmatrix}.
\end{equation}
The operation rotates the field into the principal-axis frame, applies the phase shifts, and rotates back. Expanding the product yields the explicit matrix elements:
\begin{equation}
\begin{aligned}
J_{xx} &= e^{i\phi_x}\cos^2\theta + e^{i\phi_y}\sin^2\theta,\\
J_{yy} &= e^{i\phi_x}\sin^2\theta + e^{i\phi_y}\cos^2\theta,\\
J_{yx} &= J_{xy} = \left(e^{i\phi_x}-e^{i\phi_y}\right)\sin\theta\cos\theta,
\end{aligned}
\end{equation}
where \(\theta,\phi_x,\phi_y\) abbreviate \(\theta_i^{m}(x_i,y_i),\phi_{x_i}^{m}(x_i,y_i),\phi_{y_i}^{m}(x_i,y_i)\). The three-component field after modulation on the m-th layer is
\begin{equation}
\vec{E_i^m}'(x_i,y_i)=
\begin{bmatrix}
E'_x(x_i,y_i)\\
E'_y(x_i,y_i)\\
E'_z(x_i,y_i)
\end{bmatrix}.
\end{equation}

\subsection{Full-Vector Propagation Between Diffractive Layers}
After modulation on the \(m\)-th layer, the field propagates over a distance \(d\) in free space to reach the next diffractive layer. We treat this propagation in the frequency domain using the full-vector angular spectrum method, which decomposes the transverse field into a set of plane waves. For each spatial frequency \((k_x,k_y)\), the corresponding plane-wave component advances by a phase factor \(e^{ik_z d}\).

The vector angular spectrum of the modulated field on the \(m\)-th layer is defined as
\begin{equation}
\tilde{\vec{E}}^m(k_x,k_y;0)=\mathcal{F}_{2D}\{\vec{E^m}'(x,y)\},
\end{equation}
where \(\mathcal{F}_{2D}\{\cdot\}\) denotes the two-dimensional Fourier transform. On an \(N\times N\) discrete grid with sampling intervals \(\Delta x=\Delta y=\Delta\), the spatial frequencies are given by \(k_x=2\pi f_x\), \(k_y=2\pi f_y\), and \(k=2\pi/\lambda\). The free-space dispersion relation \(k_x^2+k_y^2+k_z^2=k^2\) gives the longitudinal wavenumber:
\begin{equation}
k_z=\sqrt{k^2-k_x^2-k_y^2}.
\end{equation}
Only components satisfying \(k_x^2+k_y^2 \le k^2\) are retained, and evanescent components are discarded, following standard practice.

A key requirement for full-vector propagation is that the field satisfies the transversality condition \(\nabla\cdot\vec{E}=0\). In the plane-wave picture, this becomes \(\vec{k}\cdot\tilde{\vec{E}}=0\), meaning the spectral field vector must be orthogonal to the propagation direction \(\vec{k}=(k_x,k_y,k_z)^T\), where T denotes transpose. When \(k_x\) and \(k_y\) are nonzero, even an initially transverse field acquires a longitudinal component \(E_z\), coupling the three Cartesian components. To enforce transversality, we apply a projection operator \(\mathbf{P}\) in the frequency domain:
\begin{equation}
\mathbf{P}=\mathbf{I}-\vec{\sigma}\vec{\sigma}^{T},
\end{equation}
where the normalized wavevector $\vec{\sigma}=(\sigma_x,\sigma_y,\sigma_z)^T=\frac{1}{k}(k_x,k_y,k_z)^T$. Explicitly,
\begin{equation}
\mathbf{P}=
\begin{bmatrix}
1-\sigma_x^2 & -\sigma_x\sigma_y & -\sigma_x\sigma_z\\
-\sigma_y\sigma_x & 1-\sigma_y^2 & -\sigma_y\sigma_z\\
-\sigma_z\sigma_x & -\sigma_z\sigma_y & 1-\sigma_z^2
\end{bmatrix}.
\end{equation}
This operator removes the component parallel to \(\vec{\sigma}\) and retains only the transverse part, thereby ensuring \(\vec{k}\cdot\tilde{\vec{E}}=0\) throughout propagation. The propagated vector angular spectrum after distance \(d\) is then expressed as
\begin{equation}
\tilde{\vec{E}}(k_x,k_y;d)=\mathbf{P}(k_x,k_y) \cdot \tilde{\vec{E}}(k_x,k_y;0)\,e^{ik_z d},
\end{equation}
and the corresponding spatial-domain field on the \((m+1)\)-th layer is recovered by inverse Fourier transform:
\begin{equation}
\vec{E}^{m+1}(x,y)=\mathcal{F}^{-1}_{2D}\{\tilde{\vec{E}}(k_x,k_y;d)\}.
\end{equation}

In the dyadic Green's function formalism, the plane-wave expansion leads to a frequency-domain kernel \(\tilde{\mathbf{G}}(k_x,k_y;z)\) that maps sources to fields. The angular spectrum method, by contrast, provides a field-to-field transfer matrix \({\mathbf{P}}(k_x,k_y)e^{ik_z d}\).  In the propagating-wave regime, the two approaches differ only by a factor in the frequency domain. Here, We choose the angular spectrum method because it maps directly to how we step through the layers numerically: given the field on one layer, we compute the field on the next, while keeping the transversality condition intact. The corresponding analysis are list in supplement A.

\subsection{Combined Operator}
We now assemble the local Jones modulation and the full-vector angular spectrum propagation into a single operation that takes the field from one layer to the next. The modulation is now written as a $3\times3$ matrix acting on the three-component field:
\begin{equation}
\mathbf{J}^m(x,y)=
\begin{bmatrix}
J_{xx} & J_{yx} & 0\\
J_{xy} & J_{yy} & 0\\
0 & 0 & 1
\end{bmatrix},
\end{equation}
so the neuron output  is simply described as
\begin{equation}
\vec{E}^m_{\mathrm{out}}(x,y)=\mathbf{J}^m(x,y)\,\vec{E}^m_{\mathrm{in}}(x,y).
\end{equation}
The overall mapping from layer $m$ to layer $m+1$ can then be written as
\begin{equation}
\vec{E}^{m+1}_{\mathrm{in}}(x,y)
=
\mathcal{F}^{-1}_{2D}\!\left\{
\mathbf{P}(k_x,k_y)\cdot
\mathcal{F}_{2D}\!\left\{\vec{E}^m_{\mathrm{out}}(x,y)\right\}
\,e^{ik_z d}
\right\},
\end{equation}
where $d$ is the propagation distance between adjacent layers.

Although the modulation acts only on the transverse components directly, the propagation operator $\mathbf{P}(k_x,k_y)$ is a full $3\times3$ matrix in the frequency domain that mixes the three Cartesian components. So even if the field starts with only an $E_x$ component, after several rounds of modulation and propagation, $E_y$ and $E_z$ will generally emerge. This mixing is precisely what allows V-D$^2$NN to handle vector electromagnetic propagation and polarization coupling—a capability that scalar D$^2$NNs lack.

\section{Validation}
In this section, we validate the proposed V-D$^2$NN across three representative applications: vector beam generation, optical imaging, and multi-channel optical encryption.
\begin{figure}[!htpb]
    \centering
   \includegraphics[width=0.6\linewidth]{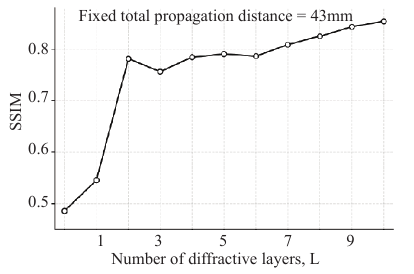}
    \caption{SSIM performance of the proposed V-D$^2$NN against the diffractive layers.}
    \label{fig:Fig2}
\end{figure}

Before the validation, we first examine how the number of diffractive layers $L$ affects imaging performance under a fixed total propagation distance of $43$ mm, with the structural similarity index (SSIM) defined in Supplementary II as the quantitative metric. As shown in Fig.~\ref{fig:Fig2}, the imaging performance does not simply improve with more layers. The SSIM increases sharply when $L$ goes from 1 to 2, but the gain becomes marginal for $L \ge 3$. This suggests that with the total propagation distance fixed, performance is not governed by layer count alone. Instead, there is a trade-off: additional layers offer more modulation capacity, but they also shorten the propagation distance between consecutive layers. Based on these observations, all subsequent validation results are obtained using the optimal configuration $L = 5$ .

\subsection{Vector Vortex Beam Generation}
Conventional diffractive deep neural networks (D$^2$NNs) are built on scalar diffraction theory, which limits their modulation to the amplitude and phase of the optical field and leaves the polarization degree of freedom largely untapped. This limitation becomes especially apparent when generating vectorial beams with complex polarization structures, such as vector vortex beams. To produce a vector vortex beam carrying a topological charge $\ell$, the usual approach involves training two independent D$^2$NNs \cite{Xiao2026}, or two separate branches in the case of R-D$^2$NNs \cite{Hu2026}, where one is for a right-circularly polarized OAM mode with charge $+\ell$ and another is for its left-circularly polarized counterpart with charge $-\ell$. These two orthogonally polarized beams must then be coherently combined through an external optical path, which demands careful alignment and phase stability. This two-path synthesis adds system complexity, insertion loss, and mode mismatch, all of which degrade the quality and practicality of the output beam.

\begin{figure}[!htpb]
    \centering
   \includegraphics[width=0.7\linewidth]{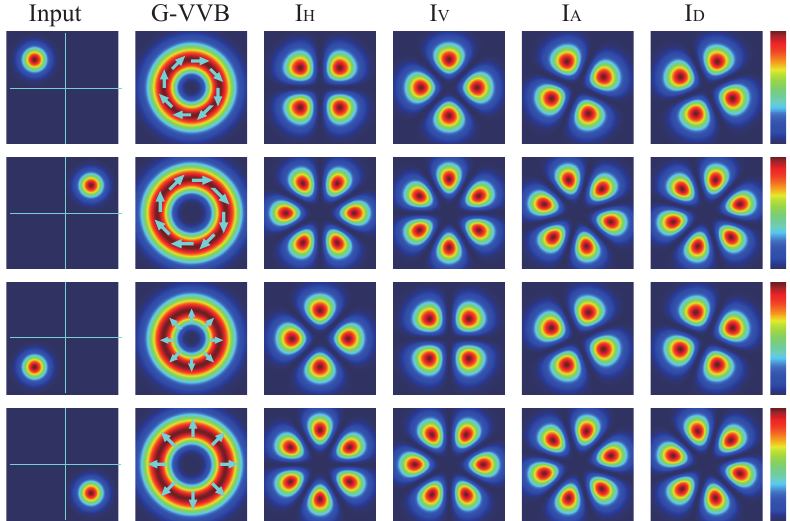}
    \caption{Generation of VVBs using the proposed V-D$^2$NN and their polarization decomposition. G-{VVB} denotes the generated VVBs, $I_H, I_V, I_A, I_D$ denote the decomposed intensities along horizontal (H), vertical (V), $135^\circ$ diagonal (A), and $45^\circ$ diagonal (D), respectively.}
    \label{fig:Fig3}
\end{figure}

Figure~\ref{fig:Fig3} illustrates the generation of vector vortex beams (VVBs) using the proposed V-D$^2$NN, in which the network learns a nonlinear mapping from the input polarization state to a prescribed VVB output. In the numerical experiments, the input plane is divided into four regions. The upper two regions are designed to generate azimuthally polarized vortex beams with two distinct topological charges, while the lower two regions are configured to produce radially polarized vortex beams. The network comprises five diffractive hidden layers, and the output plane displays the synthesized VVBs.

The incident beam is a linearly polarized Gaussian beam (horizontal polarization, wavelength $\lambda = 632.8~\mathrm{nm}$). As it propagates through the five-layer trainable vectorial network, the output intensity patterns are optimized to match the corresponding target distributions. Each diffractive layer incorporates a spatially variant Jones matrix modulator, with three trainable parameters per pixel: two phase shifts and one rotation angle. The network is trained using the Adam optimizer to minimize the complex-domain mean squared error between the output and the target field.

The results show that the desired radially and azimuthally polarized VVB states with topological charges $\ell = 2,3$ are successfully generated, confirming that the target VVBs can be obtained by controlling the input position. To verify the vector nature of the generated beams, we decompose them along the main polarization axes. The distinct patterns observed in the polarization channels $I_H$, $I_V$, $I_A$, and $I_D$ clearly confirm their vector vortex nature, and the number of petals agrees with the expected topological charge $\ell = 2,3$. While the overall VVB intensity maintains a doughnut-shaped profile, the decomposed intensities vary noticeably between states, reflecting the diversity of polarization structures accessible through the network. These results collectively demonstrate that the proposed V-D$^2$NN can generate high-fidelity vector vortex beams with well-controlled polarization and phase characteristics. Different VVB states with other topological charges can be obtained by simply redesigning the input layer layout.


\subsection{Optical Encryption}
\begin{figure}[!htpb]
    \centering
    \includegraphics[width=0.8\linewidth]{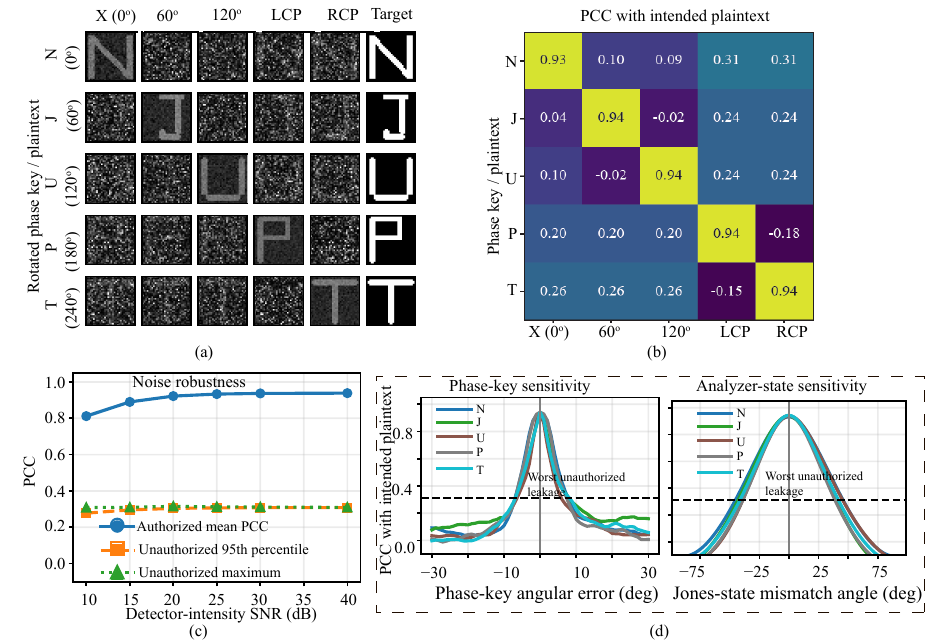}
    \caption{The demonstration of the proposed V-D$^2$NN on an optical encryption task.}
    \label{fig:Fig4}
\end{figure}

Figure~\ref{fig:Fig4} demonstrates the versatility of the proposed V-D$^2$NN in an optical encryption task. The plaintext consists of five predefined target patterns corresponding to the letters \textbf{N}, \textbf{J}, \textbf{U}, \textbf{P}, and \textbf{T}, each associated with a distinct encryption key. These keys are implemented as rotational phase masks applied at the input plane. Rather than being generated independently, all five keys are derived from a single base phase template rotated by \(0^\circ\), \(60^\circ\), \(120^\circ\), \(180^\circ\), and \(240^\circ\), respectively. Each rotation angle produces a distinct input key that introduces a unique spatial phase delay profile onto a fixed linearly polarized Gaussian beam, thereby modifying its wavefront. The resulting field propagates through the V-D$^2$NN and is modulated by the five diffractive layers, producing a polarization-resolved response at the output plane.

For selective readout of the encrypted information, the receiver is equipped with five predefined polarization analysis channels: \(X\)-polarization at \(0^\circ\), \(60^\circ\), \(120^\circ\), left-circular polarization, and right-circular polarization. Each rotational phase key is paired with a specific analysis channel, which in turn corresponds to a target plaintext pattern. Correct reconstruction is achieved only when both the input key and the analysis polarization match their designated assignments. Mismatches in either the key angle or the analysis channel, or observation through an orthogonal polarization state, yield only low-contrast, uncorrelated patterns from which the original information cannot be recovered.

The encryption performance is systematically evaluated through four complementary metrics, as summarized in Fig.~\ref{fig:Fig4}. Figure~\ref{fig:Fig4}(a) validates the selective readout capability of the proposed scheme. After training, the system establishes a well-defined one-to-one mapping among the input rotational phase keys, the output polarization analysis channels, and the target plaintext patterns. When the correct key is applied and the output is analyzed through the matching polarization channel, the corresponding letters \textbf{N}, \textbf{J}, \textbf{U}, \textbf{P}, and \textbf{T} are clearly recovered at the diagonal positions of the output array. Under mismatched key--channel conditions, the outputs appear as random speckle-like distributions with no discernible information. These results confirm that the V-D$^2$NN-based encryption framework enables effective decryption under correct key--channel pairing while simultaneously suppressing information leakage and crosstalk in incorrect polarization channels.

Figure~\ref{fig:Fig4}(b) presents the correlation matrix between the five phase keys and the five target plaintexts, quantified by the Pearson correlation coefficient (PCC) as defined in Supplement~II. The diagonal elements, corresponding to correct key--plaintext pairings, exhibit high correlation values above 0.9, while the off-diagonal elements remain near zero. This confirms that the system establishes a reliable one-to-one mapping between each key and its designated plaintext.

Figure~\ref{fig:Fig4}(c) evaluates the robustness of the encryption system against detection noise. The authorized mean PCC, the unauthorized 95th percentile, and the unauthorized maximum PCC are plotted as functions of the detector-intensity signal-to-noise ratio (SNR). The authorized mean PCC remains consistently above 0.85 across the entire SNR range, while both unauthorized metrics stay below 0.2. This confirms that the system maintains reliable decryption performance with minimal information leakage even under low-SNR conditions.

Figure~\ref{fig:Fig4}(d) examines the sensitivity of the system to phase-key angular errors and the worst-case unauthorized leakage due to polarization mismatch. The left panel shows the PCC with the intended plaintext as a function of the angular deviation of the input phase key from its nominal value. For each of the five target patterns, the PCC drops sharply as the angular error increases, with negligible correlation observed for deviations exceeding approximately \(10^\circ\). This steep roll-off indicates that the system is highly sensitive to the exact key angle, rendering brute-force or approximate-key attacks ineffective. The right panel evaluates the worst-case unauthorized leakage as a function of the Jones-state mismatch angle. The unauthorized leakage, defined as the maximum PCC observed under any incorrect polarization analysis channel, is consistently suppressed below 0.2 across all mismatch angles, with the lowest leakage occurring near orthogonal polarization states. This demonstrates that the encryption scheme is robust against unauthorized readout attempts even when the adversary employs a polarization state that partially matches the legitimate channel.

Unlike conventional encryption schemes that rely solely on single-step phase modulation or direct polarization mapping, the proposed approach simultaneously exploits three complementary mechanisms: rotational phase key encoding, vectorial propagation through birefringent elements, and multi-channel polarization-selective readout. The plaintext information is thus embedded jointly in the spatial and polarization distributions of the output field. During training, the optimization not only ensures faithful reconstruction of each plaintext in its designated channel but also actively suppresses cross-channel leakage, energy spillover into non-target channels, and correlation between incorrect keys and legitimate plaintexts. This design substantially enhances the system's information-hiding capability under incorrect observation conditions. The proposed optical encryption scheme thus establishes a joint encryption and selective decryption mechanism governed by the combined action of the phase keys, vectorial propagation, and polarization analysis channels.

\section{Comparison}
\subsection{Comparison with D$^2$NN in Handwritten Digit Classification}
\begin{figure}[!htpb]
    \centering
    \includegraphics[width=0.8\linewidth]{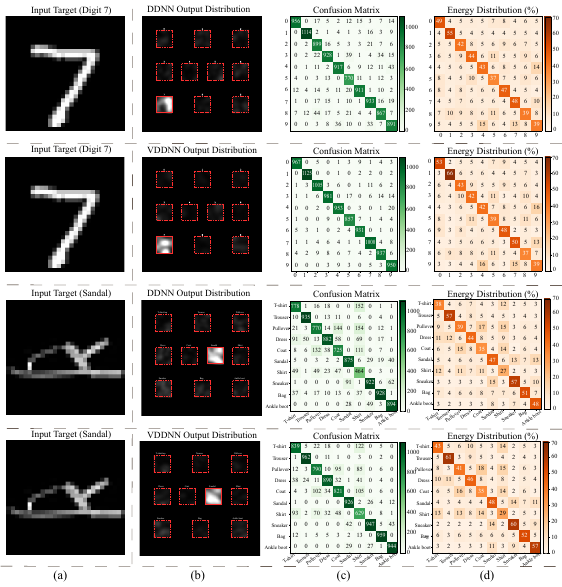}
    \caption{Classification results of the proposed V-D$^2$NN and the conventional D$^2$NN on MNIST and Fashion-MNIST.}
    \label{fig:Fig5}
\end{figure}
In this subsection, we apply the proposed V-D$^2$NN to handwritten digit and fashion item classification and compare its performance with that of the conventional D$^2$NN. Conventional D$^2$NNs are based on scalar diffraction theory and model only the spatial propagation of intensity and phase. In contrast, the V-D$^2$NN explicitly incorporates all three electric-field components ($E_x$, $E_y$, $E_z$) and enables polarization-selective modulation at each diffractive neuron. This vectorial formulation provides a more accurate physical description of light--matter interactions and is inherently better suited for tasks involving polarization-dependent or vectorial optical information.

Fig.~\ref{fig:Fig5} compares the classification performance of the proposed V-D$^2$NN and the conventional D$^2$NN on the MNIST and Fashion-MNIST datasets. Both networks successfully learn to map distinct input patterns to their corresponding detector regions. The confusion matrices for the MNIST digit ``7'' and the Fashion-MNIST item (e.g., ``sneaker'') exhibit strictly diagonal entries, confirming that all test samples are correctly classified. Notably, the V-D$^2$NN achieves a clear improvement in classification accuracy over the conventional D$^2$NN. For the MNIST dataset, the accuracy increases from $91.86\%$ (D$^2$NN) to $97.06\%$ (V-D$^2$NN); for the Fashion-MNIST dataset, the improvement is $6.53\%$. This improvement stems from the full-vector propagation model, which captures cross-polarization coupling and longitudinal field components that are neglected in scalar D$^2$NNs.

Furthermore, the output energy distributions show that the optical power for each target class is more strongly concentrated in its designated detection region, indicating better spatial energy confinement and physical interpretability. Here, the energy fraction is defined as the ratio of the energy collected within the target detection region to the total energy at the output plane. For the MNIST dataset, the energy fraction is $44.2\%$ for the D$^2$NN and $46.0\%$ for the V-D$^2$NN; for the Fashion-MNIST dataset, it is $44.43\%$ for the D$^2$NN and $47.24\%$ for the V-D$^2$NN. In addition, the results on Fashion-MNIST confirm the generalization capability of the V-D$^2$NN across more complex and diverse object categories.

\subsection{Comparison with D$^2$NN and M-D$^2$NN in Optical Imaging}
\begin{figure}[!htpb]
    \centering
    \includegraphics[width=0.8\linewidth]{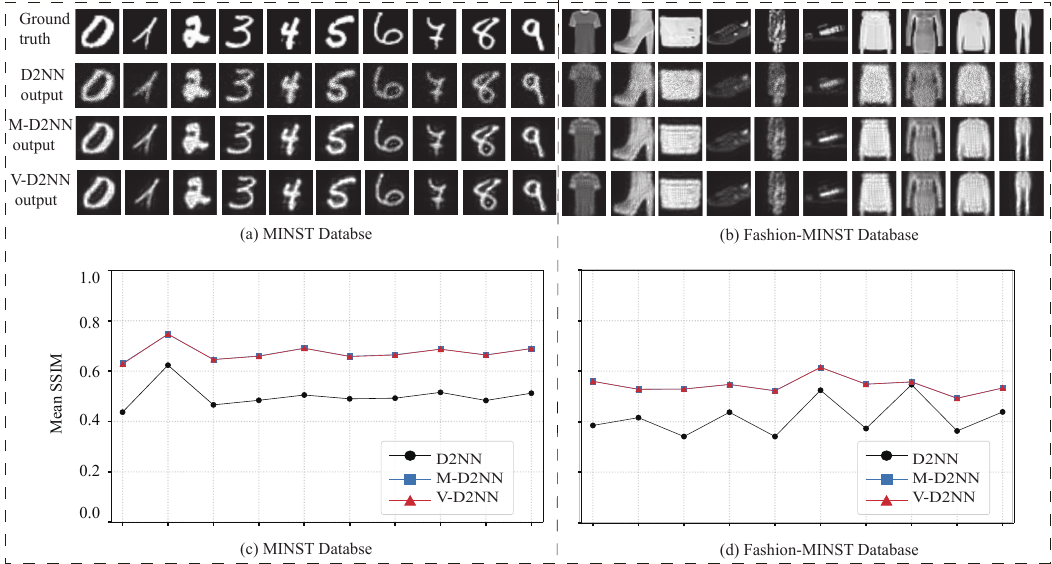}
    \caption{The proposed V-D$^2$NN used in optical imaging, together with those by D$^2$NN and M-D$^2$N.}
    \label{fig:Fig6}
\end{figure}

Figure~\ref{fig:Fig6} demonstrates the imaging capability of the proposed V-D$^2$NN on optical imaging tasks, together with the results obtained by the conventional D$^2$NN~\cite{Lin2018} and the matrix D$^2$NN (M-D$^2$NN)~\cite{Wang2024}. The network architecture and training procedure are the same as those used for vector beam generation, except that the output is now an intensity image. Importantly, the V-D$^2$NN is not trained directly on handwritten digits or fashion items. Instead, it is optimized on programmatically generated structured patterns, including white noise, band-limited noise, sinusoidal fringes, checkerboards, slanted edges, Siemens star targets, and sparse point objects. The subsequent evaluations on the MNIST and Fashion-MNIST datasets therefore serve as an independent test of the network's generalization capability.

We first evaluate the imaging performance on MNIST digits. Representative blind-test results are shown in Fig.~\ref{fig:Fig6}(a), and the mean SSIM values are summarized in Fig.~\ref{fig:Fig6}(c). For these previously unseen digit samples, the V-D$^2$NN preserves the main contours and essential stroke structures. Quantitative evaluations on the ten representative samples yield SSIM values ranging from $0.631$ to $0.747$, whereas those obtained by the conventional D$^2$NN range from $0.437$ to $0.624$. These results indicate that the V-D$^2$NN can produce stable intensity reconstructions for targets with relatively simple and well-defined boundaries.

Figure~\ref{fig:Fig6}(b) and Fig.~\ref{fig:Fig6}(d) present representative blind-test results and the corresponding mean SSIM performance on the Fashion-MNIST dataset, respectively. Compared with MNIST, Fashion-MNIST targets generally have more complex contours and larger intra-class variations, leading to more pronounced performance fluctuations across categories. For the ten representative samples shown, the SSIM achieved by the V-D$^2$NN and M-D$^2$NN ranges from $0.528$ to $0.615$, while that obtained by the conventional D$^2$NN remains below $0.525$. Overall, the V-D$^2$NN recovers the global shapes and major structural features of the fashion items, although its ability to preserve fine details and complex textures is somewhat limited. It should be noted that these values correspond to the representative samples shown in the figure, not averaged over the entire test set.

The results in Fig.~\ref{fig:Fig6} show that after framework optimization, the proposed V-D$^2$NN not only produces recognizable reconstructions for unseen digits and fashion items but also effectively retrieves structured patterns. These findings confirm that the polarization-vectorial diffractive network can perform explicit imaging functions with reasonable fidelity across different target types. Notably, since the imaging task mainly involves the transverse polarization components, the V-D$^2$NN and M-D$^2$NN exhibit similar performance, as both properly account for transverse polarization modulation, whereas the conventional D$^2$NN, which neglects polarization entirely, shows a clear degradation. The V-D$^2$NN remains advantageous in that it provides a fully vectorial description that is also valid for tasks where longitudinal components and cross-polarization coupling become non-negligible.

\subsection{Comparison with M-D$^2$NN to control $E_z$}
\begin{figure}[!htpb]
    \centering
    \includegraphics[width=0.6\linewidth]{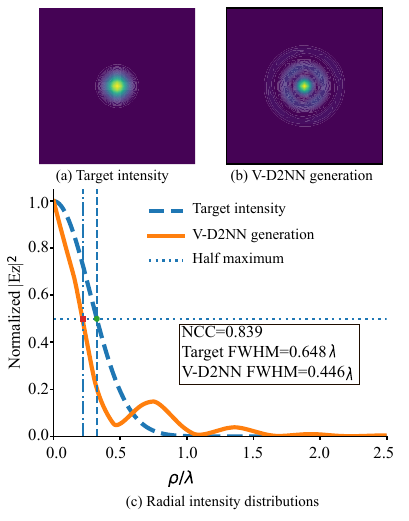}
    \caption{The normalized intensity distribution and their radial intensity distributions of the longitudinal field for the high NA.}
    \label{fig:Fig7}
\end{figure}
Existing M-D$^2$NNs use a three-parameter Jones matrix for trainable polarization modulation, enabling polarization multiplexing and multi-task processing. However, they treat $E_x$ and $E_y$ as independent scalar fields during interlayer propagation and recombine them only at the next layer, thereby neglecting the Maxwell transversality condition and the evolution of the longitudinal component $E_z$. In contrast, the proposed V-D$^2$NN retains the Jones-matrix layers but adopts full-vector angular-spectrum propagation, which enforces the vector constraint between transverse and longitudinal fields and reconstructs $E_z$ at each propagation step.

To verify the ability of the V-D$^2$NN to control $E_z$, we can construct a longitudinal-field engineering task under high Numerical Aperture(NA) conditions.
The input field is a radially polarized Gaussian beam, expressed as
\begin{equation}
E_{x,\text{in}} = G(r)\cos\phi, \qquad
E_{y,\text{in}} = G(r)\sin\phi, \qquad
E_{z,\text{in}} = 0,
\end{equation}
where $G(r) = \exp\left(-\frac{r^2}{w_c^2}\right).$
Thus, the input field contains only transverse electric-field components, and the longitudinal component is strictly set to zero, the target is defined as a circularly symmetric Gaussian distribution:
\begin{equation}
I_{z,\text{target}}(x,y) = \exp\left[-\frac{2(x^2+y^2)}{w_0^2}\right].
\end{equation}
The optimization strategy is to let the V-D$^2$NN automatically search for the transverse field distribution that can generate the target $E_z$ under the complete vector propagation constraint, while further suppressing unwanted transverse field energy in the target region through the loss function.

To simultaneously constrain the spatial shape, effective energy, and longitudinal purity of the longitudinal field, we adopt the composite loss function:
\begin{equation}
\mathcal{L} = \alpha_1 \mathcal{L}_{\text{shape}} + \alpha_2 \mathcal{L}_{\text{con}} + \alpha_3 \mathcal{L}_{\text{pur}} + \alpha_4 \mathcal{L}_{\text{floor}} + \alpha_5 \mathcal{L}_{\text{loc}},
\end{equation}
where $\alpha_1, \alpha_2, \ldots, \alpha_5$ are the weighting coefficients.  $\mathcal{L}_{\text{shape}} = 1 - C_z$, with $C_z$ the normalized correlation coefficient; $\mathcal{L}_{\text{con}} = -\log(\eta_{z,\text{core}} + \epsilon)$, with $\eta_{z,\text{core}}$ the longitudinal energy concentration ratio; $\mathcal{L}_{\text{pur}} = -\log(p_z + \epsilon)$, with $p_z$ the longitudinal purity and $\epsilon$ a small positive constant; $\mathcal{L}_{\text{floor}} = |\max(0, p_{\min} - p_z)|^2$; and $\mathcal{L}_{\text{loc}}$ suppresses undesired $E_z$ energy outside the target focal region, thereby improving the spatial localization of the longitudinal field. Detailed definitions of these terms are provided in the supplement.

Figure~\ref{fig:Fig7}(a) shows the normalized intensity distribution of the target $E_z$, Fig.~\ref{fig:Fig7}(b) shows the normalized intensity distribution of the V-D$^2$NN output $|E_z|^2$, and Fig.~\ref{fig:Fig7}(c) compares their radial intensity distributions.

The simulation results show that although the input field satisfies $E_{z,\text{in}} = 0$, a longitudinal electric field is generated at the output plane after modulation by the V-D$^2$NN and full-vector propagation. The normalized correlation coefficient between the output longitudinal field and the prescribed Gaussian target reaches $C_z = 0.839$, indicating that the V-D$^2$NN can effectively control the spatial distribution of $E_z$ according to the target. The FWHM of the output longitudinal field is $\text{FWHM}_{E_z} = 0.446\lambda$, which is narrower than the target value of $0.648\lambda$( Taking $w_0 = 0.55\lambda$, the full width at half maximum (FWHM) of the target intensity distribution is set to $\text{FWHM}_{\text{target}} = \sqrt{2\ln 2}\,w_0 \approx 0.648\lambda$.), indicating strong spatial localization of the output longitudinal field.

In terms of energy, about $\eta_{z,\text{core}} = 37.4\%$ of the total electric-field energy at the output plane is concentrated in the $E_z$ component within the target core region. Meanwhile, the longitudinal purity within the core region reaches $p_z = 67.7\%$, indicating that the longitudinal component dominates the electric field in the target focal region. At the focal center, the proportion of $E_z$ in the local total electric-field intensity approaches $100\%$, forming a focal center clearly dominated by the longitudinal electric-field component.

These results show that even when the input contains no longitudinal component, the V-D$^2$NN can exploit the physical coupling between the transverse and longitudinal fields during full-vector propagation to actively generate and shape a prescribed $E_z$ spatial distribution. It should be emphasized that the V-D$^2$NN explicitly incorporates $E_z$ into both the propagation model and the optimization objective, making the longitudinal field a physical degree of freedom that the network can directly sense, evaluate, and optimize. In contrast, the M-D$^2$NN does not explicitly describe this longitudinal-field degree of freedom during training and therefore cannot directly perform end-to-end optimization around a target $E_z$. This result highlights a key capability advantage of the V-D$^2$NN over the scalar-propagation M-D$^2$NN in longitudinal-field control, and indicates its potential value in high-NA strong focusing and other scenarios where the longitudinal electric field cannot be neglected.

\section{Conclusion}
We have proposed a vector diffractive deep neural network (V-D$^2$NN) that embeds the full-vector angular spectrum method directly into the end-to-end training pipeline. Unlike scalar or semi-vectorial models that treat polarization as an independent channel, the V-D$^2$NN explicitly accounts for the vector nature of light throughout propagation and modulation. This full-vector formulation accurately models cross-polarization coupling and vector diffraction in free-space propagation, enabling the network to exploit spin-orbit interactions and polarization-structured light fields as intrinsic computational resources.

We have demonstrated the V-D$^2$NN on several polarization-multiplexed tasks. In vector beam generation, it produces the desired vector vortex beams with controlled polarization and phase. In multi-channel optical encryption, it enables selective readout with low crosstalk and high sensitivity to the key. In polarization-dependent imaging and classification, it improves both reconstruction fidelity and recognition accuracy over the scalar D$^2$NN, while concentrating more energy in the target detection region. In longitudinal-field engineering, it actively generates and shapes a prescribed $E_z$ distribution even when the input has no longitudinal component, whereas the scalar-propagation M-D$^2$NN cannot directly optimize this degree of freedom. Across these tasks, the V-D$^2$NN consistently offers improved efficiency and design flexibility compared with scalar and semi-vectorial counterparts. An open-source training framework is also provided to support further development of vectorial diffractive optics for computing, sensing, and communications.

\begin{backmatter}
\bmsection{Funding}
National Natural Science Foundation of China (62375140, 62001249)


\bmsection{Disclosures}
\noindent The authors declare no conflicts of interest.

\bmsection{Data availability} Data underlying the results presented in this paper are not publicly available at this time but may be obtained from the authors upon reasonable request.
\end{backmatter}



\end{document}